\documentclass[a4paper,10pt]{article}
\usepackage{indentfirst}
\usepackage[colorlinks, citecolor=blue, urlcolor=blue, linkcolor=blue]{hyperref}
\usepackage{hyperref}
\usepackage{cite}
\usepackage{graphics}
\usepackage{epsfig}
\usepackage{epstopdf}
\usepackage[utf8]{inputenc}
\usepackage{url}     
\usepackage{slashed} 
\usepackage{booktabs}
\usepackage{mathrsfs}
\usepackage{epsfig}
\usepackage{dcolumn}
\usepackage{bm}
\usepackage{amsmath}
\usepackage{multirow}
\usepackage{amsmath}		
\usepackage{amssymb}		
\usepackage{caption, subcaption}
\usepackage{comment,color}

\makeatletter
\@addtoreset{equation}{section}
\makeatother
\begin{document}

\thispagestyle{empty} \vspace*{0.8cm}\hbox
to\textwidth{\vbox{\hfill\huge\sf \hfill}}
\par\noindent\rule[3mm]{\textwidth}{0.2pt}\hspace*{-\textwidth}\noindent
\rule[2.5mm]{\textwidth}{0.2pt}

\begin{center}
\LARGE\bf Properties of Heavy Higgs Bosons and Dark Matter under Current Experimental Limits in the $\mu$NMSSM
\end{center}

\begin{center}
Zhaoxia Heng$^a$\footnote{zxheng@htu.edu.cn}, Xingjuan Li$^b$\footnote{lixingjuan@stu.htu.edu.cn}, Liangliang Shang$^a$\footnote{shangliangliang@htu.edu.cn}
\end{center}

\begin{center}
\begin{footnotesize} \sl
{$^a$School of Physics, Henan Normal University, Xinxiang 453007, China \\
$^b$School of Physics and Electrical Engineering, Kashi University, Kashi 844006, China} \\
\end{footnotesize}
\end{center}

\vspace*{2mm}

\begin{center}
\begin{minipage}{15.5cm}
\parindent 20pt\footnotesize
\begin{abstract}

Searches for new particles beyond the Standard Model (SM) are an important task for the Large Hadron Collider (LHC). In this paper, we investigate the properties of the heavy non-SM Higgs bosons
in the $\mu$-term extended Next-to-Minimal Supersymmetric Standard Model ($\mu$NMSSM). We scan 
the parameter space of the $\mu$NMSSM considering the basic constraints from Higgs data, dark matter (DM) relic density, and LHC searches for sparticles. And we also consider the constraints from the LZ2022 experiment and the muon anomaly constraint at 2$\sigma$ level.
 We find that the LZ2022 experiment has a strict constraint on the parameter space of the $\mu$NMSSM, and the limits from the DM-nucleon spin-independent (SI) and spin-dependent (SD) cross-sections are complementary. 
Then we discuss the exotic decay modes of heavy Higgs bosons decaying into SM-like Higgs boson. We find that for doublet-dominated Higgs $h_3$ and $A_2$, the main exotic decay channels are $h_3\rightarrow Z A_1$, $h_3\rightarrow h_1 h_2$, $A_2\rightarrow A_1 h_1$ and $A_2\rightarrow Z h_2$, and the 
branching ratio can reach to about 23$\%$, 10$\%$, 35$\%$ and 10$\%$ respectively.
At the 13 TeV LHC, the production cross-section of $ggF\rightarrow h_3\rightarrow h_1 h_2$ and 
$ggF\rightarrow A_2\rightarrow  A_1 h_1$ can reach to about $10^{-11}$pb and $10^{-10}$pb,
respectively.

\end{abstract}
\end{minipage}
\end{center}

\begin{center}
\begin{minipage}{15.5cm}
\begin{minipage}[t]{2.3cm}{\bf Keywords:}\end{minipage}
\begin{minipage}[t]{13.1cm}
$\mu$NMSSM; heavy Higgs bosons; DM
\end{minipage}\par\vglue8pt
\end{minipage}
\end{center}

\newpage
\tableofcontents
\section{Introduction}
\label{introduction}
In July 2012, both the ATLAS and CMS collaborations at the Large Hadron Collider (LHC) 
announced a scalar with mass near 125 GeV \cite{ATLAS:2012ae,CMS:2012qbp,CMS:2012zhx}, and recently the combined measurement of the muon anomalous magnetic moment by the Fermi National Accelerator Laboratory (FNAL) \cite{Muong-2:2021ojo}
and the Brookhaven National Laboratory (BNL) \cite{Muong-2:2006rrc} showed a 4.2$\sigma$ discrepancy from 
the prediction in the Standard Model (SM).
The continuously updated experimental results provide rich information
about supersymmetry (SUSY). As an economic realization of SUSY, the Next-to-Minimal Supersymmetric Standard Model (NMSSM) \cite{Franke:1995xn,Szleper:2005jc,Ellwanger:2009wj,Franke:1995tc,Ellwanger:2009dp} has attracted more attention.
However, considering the recent experimental constraints, the parameter space of 
the NMSSM with a discrete $Z_3$-symmetry ($Z_3$-NMSSM) has been strictly constrained \cite{Cao:2022htd,Cao:2021tuh,Zhou:2021pit,Cao:2022chy}.
In order to obtain a broad parameter space that agrees with the recent experimental results, 
we extend  the $Z_3$-NMSSM by adding an explicit $\mu$-term, which is called 
$\mu$-term extended NMSSM ($\mu$NMSSM) \cite{Cao:2022chy,Cao:2021ljw}.
Compared with $Z_3$-NMSSM, the $\mu$NMSSM can easily explain
the discrepancy of the muon anomalous magnetic moment in a broad parameter space, and 
meanwhile, coincide with the experimental results in dark matter (DM) and Higgs physics, and also the LHC searches for sparticles \cite{Cao:2021tuh,Cao:2022ovk}. In addition, the $\mu$NMSSM is free from the tadpole problem and domain-wall problem in the $Z_3$-NMSSM.

Since the discovery of a 125 GeV Standard Model (SM)-like Higgs boson at the LHC, the search for non-SM Higgs bosons has become even more pressing. In the $\mu$NMSSM, the lightest or 
next-to-lightest CP-even Higgs boson can be regarded as the SM-like Higgs boson. 
In addition to the SM-like Higgs boson ($h_1$ or $h_2$), the
$\mu$NMSSM predicts another two CP-even neutral Higgs bosons ($h_1$/$h_2$ and $h_3$), two CP-odd neutral Higgs bosons ($A_1$ and $A_2$) and a pair of charged Higgs bosons ($H^{\pm}$). In this paper, we explore the discovery potential for the non-SM heavier Higgs bosons $h_3$ and $A_2$ in the $\mu$NMSSM at the LHC.

At present, besides the conventional search channels for heavy Higgs focusing on the decay modes into 
pairs of SM particles, the heavy Higgs exotic decay modes in the $\mu$NMSSM are kinematically open.
The heavy neutral Higgs bosons can have sizable
branching ratio into two lighter neutral Higgs bosons, or to a lighter neutral Higgs boson and one $Z$ boson. The relevant searches have been carried out at the LHC \cite{CMS:2019qcx,ATLAS:2017xel,ATLAS:2022enb,CMS:2015flt,CMS:2019kca,CMS:2019ogx,ATLAS:2020gxx,CMS:2021yci}. Ref.\cite{Ellwanger:2022jtd} has presented benchmark
planes with cross-sections via gluon fusion for the exotic decay channels of heavy Higgs bosons in the NMSSM. And some discussions about
the heavy Higgs exotic decays have also been conducted in the 
Two-Higgs-Doublet Model (2HDM) \cite{Kling:2018xud,Kling:2022jcd}.
However, there have been no relevant discussions regarding the properties of the heavy Higgs boson in the $\mu$NMSSM. Therefore, our study aims to investigate the properties of the heavier CP-even Higgs boson $h_3$ and CP-odd Higgs boson $A_2$ in the $\mu$NMSSM. We focus on the searches for heavy Higgs bosons in final states with two lighter scalars, or one light scalar and a $Z$ boson.

The outline of this paper is as follows: in Section 2, we briefly describe the relevant theoretical preliminaries of $\mu$NMSSM including the Higgs sector, the neutralino sector, and the DM-nucleon scattering cross-section. In Section 3, we give the numerical results considering the constraints of DM from the LZ experiment, and investigate the properties of heavy Higgs bosons. In Section 4, we list the summary of this paper.

\section{\label{sec:theo}Theoretical preliminaries}

\subsection{\label{sec:theo}The basics of the $\mu$NMSSM}
To solve the problem in the Minimal Supersymmetric Standard Model (MSSM), such as the 
$\mu$ problem, the NMSSM is introduced. 
The NMSSM consisits of two Higgs doublet superfields $\widehat{H}_{u}$ and $\widehat{H}_{d}$, 
and one singlet chiral superfield $\widehat{S}$. After the electroweak symmetry
breaking, the Higgs fields acquire the vacuum expected values (vevs), i.e. $<H_u>=v_u$, $<H_d>=v_d$, $<S>=v_s$,  and $v=\sqrt{{v_u}^2+{v_d}^2}$,  $\tan\beta=\frac{v_u}{v_d}$. 
The Higgs fields in the NMSSM can be written as follows\cite{Ellwanger:2009dp},
\begin{equation}
	\begin{gathered}
		\widehat{H}_u=\left(\begin{array}{c}
			H_{u}^{+} \\
			v_{u}+\frac{1}{\sqrt{2}}\left(\phi_{u}+i\varphi_{u}\right)
		\end{array}\right), \quad {\widehat{H}_d}=\left(\begin{array}{c}
			v_{d}+\frac{1}{\sqrt{2}}\left(\phi_{d}+i\varphi_{d}\right) \\
			H_{d}^{-}
		\end{array}\right),~~
		\widehat{S}={v_s}+\frac{1}{\sqrt{2}}\left(\phi_{s}+i\varphi_{s}\right)
	\end{gathered}
\end{equation}
where $\phi_{u}$, $\phi_{d}$ and $\phi_{s}$ denote the neutral CP-even Higgs fields,
$\varphi_u$, $\varphi_d$ and $\varphi_{s}$ denote the neutral CP-odd Higgs fields, and
$H_{u}^{+}$, $H_{d}^{-}$ denote the charged Higgs fields. 

The general form of the superpotential in the NMSSM can be given by\cite{Ellwanger:2009dp,Abel:1996cr,Maniatis:2009re},
\begin{equation}
	W_{\text {NMSSM}}=W_{\text {Yukawa}}+(\mu+\lambda \widehat{S}) 
    \widehat{H}_{u} \cdot \widehat{H}_{d}+\xi_{F} \widehat{S}+\frac{1}{2} \mu^{\prime} \widehat{S}^{2}+\frac{\kappa}{3} \widehat{S}^{3}
\end{equation}
where the term $W_{\text {Yukawa}}$ is the same as that of the MSSM, $\mu$ and $\mu^{\prime}$ are bilinear mass coefficients, $\lambda$ and $\kappa$ are dimensionless coupling coefficients, $\xi_{F}$ is the supersymmetric tadpole term of mass square dimension. And the parameters $\mu$, $\mu^{\prime}$ and $\xi_{F}$ can
be used to solve the tadpole problem and domain-wall problem in the  
$Z_3$-symmetry NMSSM\cite{Panagiotakopoulos:1998yw,Ferrara:2010in,Hollik:2018yek,Hollik:2020plc,Kolda:1998rm}. 

In this work, we consider a specific scenario in which the parameters
$\mu^{\prime}$ and $\xi_{F}$ in Eq.(2.2) are equal to 0. This special scenario can be called the $\mu$-term extended NMSSM ($\mu$NMSSM). The superpotential and the corresponding 
soft breaking Lagrangian can be written as follows \cite{Hollik:2018yek,Hollik:2020plc}
\begin{equation}
	W_{\mu \text{NMSSM}}=W_{\text {Yukawa}}+(\mu+\lambda \widehat{S}) 
    \widehat{H}_{u} \cdot \widehat{H}_{d}+\frac{\kappa}{3} \widehat{S}^{3},
\end{equation}
\begin{equation}
	\begin{aligned}
		\qquad -\mathcal{L}_{\mathrm{soft}} &=\left[A_{\lambda} \lambda S H_{u} \cdot H_{d}+\frac{1}{3} A_{\kappa} \kappa S^{3}+ B_{\mu}\mu H_{u} \cdot H_{d}+h . c .\right] +m_{H_{u}}^{2}\left|H_{u}\right|^{2}+m_{H_{d}}^{2}\left|H_{d}\right|^{2}+m_{S}^{2}|S|^{2},
	\end{aligned}
\end{equation}
where $H_u$, $H_d$ and $S$ are the scalar parts of the superfields $\widehat{H}_u$,  $\widehat{H}_d$ and $\widehat{S}$, respectively.
By solving the minimization equation of the scalar potential, the soft breaking mass
parameters $m_{H_{u}}^{2}$, $m_{H_{d}}^{2}$, $m_S^{2}$ can be expressed in terms of the vacuum expected values of the scalar field. To simplify the calculation we set $B_{\mu}$
to be 0. Therefore, the Higgs sector is partially determined by the following parameters: 
\begin{equation}
   \tan\beta, ~~\mu_{\rm eff}=\lambda v_{s}/\sqrt{2},~~ \lambda, ~~\kappa, ~~A_{\lambda},~~ A_{\kappa}, ~~\mu.  
\end{equation}


For convenience, we define $H_{\rm SM}\equiv \sin\beta \text{Re}(H_{u}^0)+\cos\beta \text{Re}(H_{d}^0)$, $H_{\rm NSM}\equiv \cos\beta \text{Re}(H_{u}^0)-\sin\beta \text{Re}(H_{d}^0)$, and $A_{\rm NSM}\equiv \cos\beta \text{Im}(H_{u}^0)-\sin\beta \text{Im}(H_{d}^0)$, where $H_{\rm SM}$ is the SM Higgs field and its vev is $v/\sqrt{2}$, $H_{\rm NSM}$ is the other CP-even doublet Higgs field and its vev is zero, and $A_{\rm NSM}$ corresponds to the CP-odd Higgs boson in the MSSM\cite{Miller:2003ay,Cao:2012fz}. 
In the basis $(H_{\rm NSM},H_{\rm SM},\text{Re}(S))$, the elements of the CP-even Higgs mass symmetric matrix $M_{S}^{2}$ can be written as
\begin{equation}
	\begin{aligned}
		M_{S, 11}^{2} &=\frac{2\mu_{eff}(\lambda A_\lambda+\kappa \mu_{eff})}{\lambda \sin2\beta}+\frac{1}{2}(2m_{Z}^2-\lambda^2v^2)\sin{^2}2\beta, \\
		M_{S, 12}^{2} &=-\frac{1}{4}(2m_{Z}^2-\lambda^2v^2)\sin4\beta, \\
		M_{S, 13}^{2} &= -\frac{1}{\sqrt2}(\lambda A_{\lambda}+2\kappa \mu_{eff})v \cos2\beta,\\
		M_{S, 22}^{2} &=m_{Z}^2\cos{^2}2\beta+\frac{1}{2}\lambda^2v^2\sin{^2}2\beta,\\
		M_{S, 23}^{2} &=\frac{v}{\sqrt2}[2\lambda(\mu_{eff}+\mu)-(\lambda A_{\lambda}+2\kappa \mu_{eff})\sin2\beta],\\
		M_{S, 33}^{2} &=\frac{\lambda A_{\lambda}\sin2\beta}{4\mu_{eff}}\lambda v^2+\frac{\mu_{eff}}{\lambda}(\kappa A_{\kappa}+\frac{4\kappa^2\mu_{eff}}{\lambda})-\frac{\mu}{2\mu_{eff}}\lambda^2v^2.
	\end{aligned}
\end{equation}
And the elements of the CP-odd Higgs mass symmetric matrix $M_{P}^{2}$ under the basis $(A_{NSM},\text{Im}(S))$ is given by
\begin{equation}
	\begin{aligned}
		&M_{P, 11}^{2}=\frac{2\mu_{eff}(\lambda A_{\lambda}+\kappa\mu_{eff})}{\lambda \sin2\beta},\\
		&M_{P, 22}^{2}=\frac{(\lambda A_{\lambda}+4\kappa\mu_{eff})\sin{2\beta}}{4\mu_{eff}}\lambda v^2-\frac{3\kappa A_{\kappa}\mu_{eff}}{\lambda}-\frac{\mu}{2\mu_{eff}}\lambda^2 v^2,\\
		&M_{P, 12}^{2}=\frac{v}{\sqrt2}(\lambda A_{\lambda}-2\kappa\mu_{eff}).
	\end{aligned}
\end{equation}
By diagonalizing $M_{S}^2$ and $M_{P}^2$ using the unitary matrix $V$ and $U$,
we can obtain the CP-even Higgs mass eigenstate $h_{i}(i=1,2,3)$ 
with $m_{h_1}<m_{h_2}<m_{h_3}$, 
and CP-odd Higgs mass eigenstate $A_i(i=1,2)$ with $m_{A_1}<m_{A_2}$, respectively \cite{Hollik:2020plc,King:2012tr,Wang:2015omi}.
\begin{equation}
    \begin{aligned}
        &h_{i}=V_{h_{i}}^{\rm NSM}H_{\rm NSM}+V_{h_{i}}^{\rm SM}H_{\rm SM}+V_{h_{i}}^{S}\text{Re}(S),\\
        &A_{i}=U_{A_{i}}^{\rm NSM}A_{\rm NSM}+U_{A_{i}}^{S}\text{Im}(S).
    \end{aligned}
\end{equation}
Each of the three CP-even Higgs bosons  $h_{i}(i=1,2,3)$ can be either SM-like ($h$), or
$H_{\rm NSM}$ dominant ($H$), or singlet dominant ($h_s$). Likewise,
each of the two CP-odd Higgs bosons $A_i$ can be either singlet dominant ($A_s$),
or $H_{\rm NSM}$ dominant ($A_H$).
 
The mass eigenstate of charged Higgs bosons is 
$H^{\pm}=\cos{\beta}H^{\pm}_{u}+\sin{\beta}H^{\pm}_{d}$,
and their masses can be written as
\begin{equation}
     \begin{aligned}
         &m_{H^{\pm}}^2=\frac{2\mu_{eff}(\lambda A_{\lambda}+\kappa\mu_{eff})}{\lambda\sin{2\beta}}+m_{W}^2-\lambda^2v^2.
     \end{aligned}
\end{equation}

For neutralino sector, the neutralino mass eigenstate in the basis of $
\psi^{0}=\left(-i \tilde{B}^{0},-i \tilde{W}^{0}, \tilde{H}_{d}^{0}, \tilde{H}_{u}^{0}, \tilde{S}^{0}\right)
$ is
\begin{equation}
	M_{\tilde{N}}=\left(\begin{array}{ccccc}
		M_{1} & 0 & -m_{Z} \sin \theta_{\mathrm{w}} \cos \beta & m_{Z} \sin \theta_{\mathrm{w}} \sin \beta & 0 \\
		0 & M_{2} & m_{Z} \cos \theta_{\mathrm{w}} \cos \beta & -m_{Z} \cos \theta_{\mathrm{w}} \sin \beta & 0 \\
		-m_{Z} \sin \theta_{\mathrm{w}} \cos \beta & m_{Z} \cos \theta_{\mathrm{w}} \cos \beta & 0 & -\mu_{\rm tot} & -\frac{1}{\sqrt{2}}\lambda v \sin \beta \\
		m_{Z} \sin \theta_{\mathrm{w}} \sin \beta & -m_{Z} \cos \theta_{\mathrm{w}} \sin \beta & -\mu_{\rm tot} & 0 & -\frac{1}{\sqrt{2}}\lambda v \cos \beta \\
		0 & 0 & -\frac{1}{\sqrt{2}}\lambda v \sin \beta & -\frac{1}{\sqrt{2}}\lambda v \cos \beta  & 2 \frac{\kappa}{\lambda} \mu_{\rm eff}
	\end{array}\right),
\end{equation}
where $\mu_{\rm tot}\equiv\mu+\mu_{\rm eff}$, and $M_{1}$, $M_2$ are Bino and Wino soft breaking masses, respectively.
After diagonalizing the mass matrix $M_{\tilde{N}}$ by rotation matrix $N$, we can obtain the neutralino mass eigenstate $\tilde{\chi}_i^0$(i=1,2,3,4,5) labeled in
mass-ascending order, which can be expressed as 
\begin{equation}
\tilde{\chi}_i^0=N_{ij}{\psi_j^0}(j=1,2,3,4,5).
\end{equation}
Assuming the lightest neutralino $\tilde{\chi}_1^0$ is the LSP, which can be considered as an ideal candidate for DM. Evidently, $N_{11}^2$, $N_{12}^2$, $N_{13}^2+N_{14}^2$ and $N_{15}^2$ 
denote the Bino, Wino, Higgsino and Singlino fractions in $\tilde{\chi}_1^0$, respectively.
Different from the case in the $Z_3$-NMSSM, 2|$\kappa$| may be much larger than $\lambda$
in obtaining Singlino-dominated DM.

\subsection{\label{sec:heavy}The Heavy Higgs bosons}
In this work we require the lightest CP-even Higgs boson $h_1$ is SM-like, and investigate 
the properties of the heavy Higgs bosons $h_3$ and $A_2$. 
At the LHC, the heavy Higgs boson $H$ ($h_3$ or $A_2$) is mainly produced through 
gluon-gluon fusion (ggF),  and the production cross-section can be obtained by
 \begin{equation}
    \frac{\sigma(ggF\rightarrow H)}{\sigma(ggF\rightarrow h_{SM})}=\left|C_{ggF}^H\right|^2,
\end{equation}
where $h_{\rm SM}$ denotes the SM-like Higgs boson, and $C_{ggF}^H$ is the reduced coupling 
coefficient relative to the prediction in the SM. In the $\mu$NMSSM, the exotic decay modes of heavy Higgs bosons are open and heavy Higgs bosons 
$h_3$ and $A_2$ can have sizable branching ratio into two lighter Higgs bosons, 
e.g., $h_3\to h_1h_2$, $A_2\to A_1h_1$, which can be called Higgs-to-Higgs decays.
In addition, heavy Higgs bosons $h_3$ and $A_2$ may decay into one light Higgs boson and a $Z$ boson, e.g., $h_3\to A_1Z$, $A_2\to h_1Z$. The branching ratio
 of the Higgs-to-Higgs decays depends on trilinear Higgs couplings. For the typical case with $v_s$, $A_\lambda\gg v_u, v_d\approx M_Z$, the relevant trilinear Higgs couplings relative to
 Higgs-to-Higgs decays can be expressed by (neglecting contributions 
 of $\mathcal{O}(M_Z)$) \cite{Ellwanger:2009dp,Ellwanger:2022jtd}
\begin{equation}
    \begin{aligned}
  & (1) \sim H_{\rm NSM}H_{\rm SM}\text{Re}(S): \ \ \ \ \  \ \ \  -\frac{\lambda}{\sqrt{2}}(2\kappa v_s+A_{\lambda}),\\ 
   &(2) \sim A_{\rm NSM}H_{\rm SM}\text{Im}(S): \ \ \ \ \ \  \sqrt{2}\lambda\sin{\beta}\cos{\beta}(-2\kappa v_s+A_{\lambda}).
    \end{aligned}     \label{trilinear coupling}
\end{equation}
For the decays $h_i\rightarrow A_j+Z $ and $A_j \rightarrow h_i +Z$, the relevant couplings are:
\begin{equation}
    h_i(p)A_j(p^{\prime})Z_{\mu}: \ \ \ \ \ -igV_{h_{i}}^{\rm NSM}U_{A_{i}}^{\rm NSM}(p-p^{\prime})_{\mu}
\end{equation}
where $V_{h_{i}}^{\rm NSM}$ is the $H_{\rm NSM}$ component of the physical state $h_i$, and $U_{A_{i}}^{\rm NSM}$ is the $A_{\rm NSM}$ component of the physical state $A_j$.

\subsection{\label{sec:theo}The anomalous magnetic moment of the muon in the $\mu$NMSSM}
The recent measurement of the muon anomalous magnetic moment $a_{\mu}^{exp}$ by the FNAL has been updated, and its value is \cite{Muong-2:2021ojo}:
\begin{equation}
    a_{\mu}^{exp}(\text{FNAL})=116592040(54)\times10^{-11}.
\end{equation}
The result $a_{\mu}^{exp}(\text{FNAL})$ is in full agreement with the BNL E821 result $a_{\mu}^{exp}(\text{BNL})$\cite{Muong-2:2006rrc}
\begin{equation}
    a_{\mu}^{exp}(\text{BNL})=116592080(63)\times10^{-11},
\end{equation}
And the combined experimental average $a_{\mu}^{exp}$ is\cite{Aoyama:2020ynm,Aoyama:2012wk,Aoyama:2019ryr,Czarnecki:2002nt,Gnendiger:2013pva,Keshavarzi:2018mgv,Stoffer:2019pgd,Colangelo:2018mtw,Davier:2017zfy,Keshavarzi:2019abf,Hoferichter:2019mqg,Kurz:2014wya,Melnikov:2003xd,Masjuan:2017tvw,Hoferichter:2018kwz,Gerardin:2019vio,Colangelo:2019uex,Blum:2014oka}
\begin{equation}
    a_{\mu}^{exp}=116592061(41)\times10^{-11}.
\end{equation}

In SUSY, the contributions to $a_{\mu}^{\rm SUSY}$ mainly originate from the loops mediated by a smuon and a neutralino or a chargino and a muon-type sneutrino\cite{Li:2021xmw,Li:2021pnt,Du:2017str,Wang:2018vxp,PhysRevD.98.055015,Yang:2018guw,Li:2021koa,Wang:2021bcx,Aboubrahim:2021myl,Wang:2021lwi,Zheng:2021wnu}.
In the $\mu$NMSSM, the one-loop contributions to $a_{\mu}^{\rm SUSY}$ can be written as\cite{Domingo:2008bb}
\begin{equation}
\begin{gathered} a_{\mu}^{SUSY}=a_{\mu}^{\tilde{\chi}^0\tilde{\mu}}+a_{\mu}^{\tilde{\chi}^{\pm}\tilde{\nu}},\\
		a_{\mu}^{\tilde{\chi}^{0} \tilde{\mu}}=\frac{m_{\mu}}{16 \pi^{2}} \sum_{i, l}\left\{-\frac{m_{\mu}}{12 m_{\tilde{\mu}_{l}}^{2}}\left(\left|n_{i l}^{\mathrm{L}}\right|^{2}+\left|n_{i l}^{\mathrm{R}}\right|^{2}\right) F_{1}^{\mathrm{N}}\left(x_{i l}\right)+\frac{m_{\tilde{\chi}_{i}^{0}}}{3 m_{\tilde{\mu}_{l}}^{2}} \operatorname{Re}\left(n_{i l}^{\mathrm{L}} n_{i l}^{\mathrm{R}}\right) F_{2}^{\mathrm{N}}\left(x_{i l}\right)\right\}, \\
		a_{\mu}^{\tilde{\chi}^{\pm} \tilde{\nu}}=\frac{m_{\mu}}{16 \pi^{2}} \sum_{k}\left\{\frac{m_{\mu}}{12 m_{\tilde{\nu}_{\mu}}^{2}}\left(\left|c_{k}^{\mathrm{L}}\right|^{2}+\left|c_{k}^{\mathrm{R}}\right|^{2}\right) F_{1}^{\mathrm{C}}\left(x_{k}\right)+\frac{2 m_{\tilde{\chi}_{k}^{\pm}}}{3 m_{\tilde{\nu}_{\mu}}^{2}} \operatorname{Re}\left(c_{k}^{\mathrm{L}} c_{k}^{\mathrm{R}}\right) F_{2}^{\mathrm{C}}\left(x_{k}\right)\right\},
		\end{gathered}
\end{equation}
where $i=1,2,3,4,5$, $j=1,2$, $l=1,2$ represent the neutralino, chargino and smuon index, respectively. And 
\begin{equation}
	\begin{gathered}
		n_{i l}^{\mathrm{L}}=\frac{1}{\sqrt{2}}\left(g_{2} N_{i 2}+g_{1} N_{i 1}\right) X_{l 1}^{*}-y_{\mu} N_{i 3} X_{l 2}^{*}, \quad n_{i l}^{\mathrm{R}}=\sqrt{2} g_{1} N_{i 1} X_{l 2}+y_{\mu} N_{i 3} X_{l 1}, \\
		c_{k}^{\mathrm{L}}=-g_{2} V_{k 1}^{\mathrm{c}}, \quad c_{k}^{\mathrm{R}}=y_{\mu} U_{k 2}^{\mathrm{c}}, \quad
  x_{il}=m^2_{\tilde{\chi}^0_i}/m^2_{\tilde{\mu}_l}, \quad
  x_{k}=m^2_{\tilde{\chi}^{\pm}_k}/m^2_{\tilde{\nu}_{\mu}},
	\end{gathered}
\end{equation}
where $X$ denotes the smuon mass rotation matrices, $U^C$ and $V^C$ denote the chargino mass rotation matrix. 
$F_1(x)$ and $F_2(x)$ are loop functions of the kinematic variables $x_{il}$ and $x_{k}$, and their expressions are written as \cite{Athron:2015rva,Endo:2021zal}

\begin{equation}
	\begin{aligned}
		&F_{1}^{N}(x)=\frac{2}{(1-x)^{4}}\left[1-6 x+3 x^{2}+2 x^{3}-6 x^{2} \ln x\right], \\
		&F_{2}^{N}(x)=\frac{3}{(1-x)^{3}}\left[1-x^{2}+2 x \ln x\right], \\
		&F_{1}^{C}(x)=\frac{2}{(1-x)^{4}}\left[2+3 x-6 x^{2}+x^{3}+6 x \ln x\right], \\
		&F_{2}^{C}(x)=-\frac{3}{2(1-x)^{3}}\left[3-4 x+x^{2}+2 \ln x\right].
	\end{aligned}
\end{equation}

For the scenario with mass-degenerate sparticles, the relationship $F_{1}^{N}(1)=F_{2}^{N}(1)=F_{1}^{C}(1)=F_{2}^{C}(1)=1$ holds. 

\subsection{\label{sec:theo}DM-nucleon scattering cross-section}
In the $\mu$NMSSM, the lightest neutralino $\tilde{\chi}_1^0$ as the LSP can be considered as a DM candidate\cite{Almarashi:2022iol,Wang:2023suf}. The Higgsino fraction of $\tilde{\chi}_1^0$ plays an important role in elastic scattering between $\tilde{\chi}_1^0$ and nucleon. In the scenario with massive squarks, 
the spin-dependent (SD) scattering of $\tilde{\chi}_1^0$ with nucleons is mediated by exchanging
a $Z$ boson, and the scattering cross-section is approximated given by
\cite{Badziak:2015exr,Badziak:2017uto}:
\begin{equation}
    \sigma^{\rm SD}_{\tilde{\chi}_1^0-N}\simeq C_N \times \left(\frac{N_{13}^2-N_{14}^2}{0.1}\right)^2,
\end{equation}
where $N = p(n)$ denoting protons (neutrons) and $C_p\simeq4\times10^{-4}$pb,
$C_n\simeq3.1$pb, and 
\begin{equation}
    N_{13}^2-N_{14}^2=(\frac{\lambda v}{\sqrt{2}\mu_{tot}})^2\frac{N^2_{15}\cos{2\beta}}{1-(m_{\tilde{\chi}_1^0}/\mu_{tot})^2}.
\end{equation}

The spin-independent (SI) scattering of $\tilde{\chi}_1^0$ with nucleons is mainly produced by exchanging CP-even Higgs bosons through the t-channel, and the cross-section is as follows \cite{Badziak:2015nrb,Pierce:2013rda,Cao:2021ljw}
\begin{equation}
    \sigma^{SI}_{\tilde{\chi}_1^0-N}=\frac{4\tilde{\mu}_R^2}{\pi}\left|f^{(N)}\right|^2,
\end{equation}
where $\tilde{\mu}_R\equiv m_Nm_{\tilde{\chi}_1^0}/(m_N+m_{\tilde{\chi}_1^0})$ denoting the reduced mass of the DM-nucleon system. The expressions of the effective couplings $f^{(N)}$ are 
\begin{equation}
    f^{(N)}= \sum_{h_i={h,H,h_S}} f_{h_i}^{(N)}=\sum_{h_i={h,H,h_S}}\frac{C_{\tilde{\chi}_1^0\tilde{\chi}_1^0h_i}C_{NNh_i}}{2m_{h_i}^2},
\end{equation}
with $C_{NNh_i}$ being the coupling coefficient between CP-even Higgs bosons and nucleon,
\begin{equation}
    C_{NNh_i}=-\frac{m_N}{v}\left[F_d^{(N)}(V_{h_i}^{\rm SM}-\tan{\beta}V_{h_i}^{\rm NSM})+F_u^{(N)}\left(V_{h_i}^{\rm SM}+\frac{1}{\tan{\beta}}V_{h_i}^{\rm NSM}\right)\right],
\end{equation}
where $F^{(N)}_d=f_d^{(N)}+f_s^{(N)}+\frac{2}{27}f_G^{(N)}$ and $F_u^{(N)}=f_u^{(N)}+\frac{4}{27}f_G^{(N)}$. The form factors $f_q^{(N)}=m_N^{-1}\left<N|m_qq\overline{q}|N\right>$ ($q=u,d,s$)
denote the normalized light quarks contribution to the nucleon mass, and 
$f_G^{(N)}=1-\sum_{q=u,d,s}f_q^{(N)}$ represents other heavy quarks mass fraction in the nucleon.

\section{\label{sec:theo}Numerical Results}
The parameter space of the $\mu$NMSSM has been scanned by EasyScan\_HEP  \cite{Shang:2023gfy} with the Metropolis-Hastings algorithm:
\begin{equation}
	\begin{array}{c}
	\left|M_{1}\right| \leq 1.5 \mathrm{TeV}, \quad 100 \mathrm{GeV} \leq M_{2} \leq 1.5 \mathrm{TeV}, \\
		0 \leq \lambda \leq 0.75, \quad|\kappa| \leq 0.75, \quad 1 \leq \tan \beta \leq 60, \quad 2 \mathrm{TeV} \leq \left|A_{t}\right| \leq 5 \mathrm{TeV}, \\
		10 \mathrm{GeV} \leq \mu \leq 1 \mathrm{TeV}, \quad 100 \mathrm{GeV} \leq \mu_{\mathrm{tot}} \leq 1 \mathrm{TeV}, \quad\left|A_{\kappa}\right| \leq 700 \mathrm{GeV}, \\
		100 \mathrm{GeV} \leq m_{\tilde{\mu}_L} \leq 1 \mathrm{TeV}, \quad 100 \mathrm{GeV} \leq m_{\tilde{\mu}_E} \leq 1 \mathrm{TeV}.
	\end{array}
\end{equation}

For other supersymmetric parameters, we fix them to be 2 TeV. We use the package
SARAH-4.14.3 \cite{Staub:2008uz,Staub:2012pb,Staub:2013tta,Staub:2015kfa} to generate the model files in the
$\mu$NMSSM, use the program SPheno-4.0.4 \cite{Porod:2003um,Porod:2011nf} to obtain the 
particle spectrum, and use the package MicrOMEGAs-5.2.13 \cite{Belanger:2014vza,Belanger:2001fz,Belanger:2013oya,Belanger:2010pz,Belanger:2010gh,Belanger:2008sj,Belanger:2007zz,Belanger:2006is,Belanger:2004yn,Belanger:2002nx} to calculate the DM observables.

To be specific, we require the samples to satisfy the following basic constraints:

1. The lightest CP-even Higgs boson $h_1$ to be SM-like, and its mass should be between 121 GeV and 129 GeV. We utilize the code HiggsSignals-2.2.3 \cite{Bechtle:2014ewa} to fit the properties of the SM-like Higgs boson to LHC Higgs data, and utilize the code HiggsBounds-5.3.2 \cite{Bechtle:2015pma} to implement the constraints from the direct
search for extra Higgs bosons at the LEP and Tevatron. 

2. We assume the lightest neutralino is one of the DM candidates, so when comparing the dark matter scattering cross-section below with the experimental limit, we need the DM relic density $\Omega h^2 \textless 0.120$\cite{Bagnaschi:2017tru}. The SI and SD DM cross-sections should be scaled by a factor $\Omega h^2/0.120$. 

3. We consider the constraints from the direct detection experiments for sparticles at the LHC, and use SModelS-1.2.3\cite{Khosa:2020zar,MahdiAltakach:2023bdn} to decompose the spectrum including these processes:
\begin{equation}
    \begin{aligned}
    &pp\rightarrow \tilde{\chi}_2^0 \tilde{\chi}_1^{\pm},\\
    &pp\rightarrow \tilde{\chi}_i^0 \tilde{\chi}_j^0,\ \  i=2,3,4,5;\ \ j=2,3,4,5,\\
    &pp\rightarrow \tilde{\chi}_i^{\pm}\tilde{\chi}_j^{\mp},\ \ i=1,2;\ \ j=1,2.
    \end{aligned}
\end{equation}
The cross-sections of these processes at $\sqrt{s}$ = 13 TeV are calculated at the leading order by MadGraph-aMC@NLO-3.1.0\cite{Alwall:2011uj,Conte:2012fm}, and the next-to-leading order generated by an average $K$-factor $1.5$. In the following discussions, all the surviving samples satisfy these basic constraints.

\subsection{Properties of Dark Matter}
We project all the surviving samples from the scan onto a two-dimensional diagram, as shown below. The surviving samples are divided into three categories by three different colors: the  purple samples satisfy the basic constraints mentioned above, the yellow samples  satisfy not only the basic constraints but also the muon anomaly constraint within 2$\sigma$ level, and the red samples satisfy the basic constraints, the muon anomaly constraint  within 2$\sigma$ level and also the LZ experiment constraint in the 2022 year (LZ2022) \cite{LUX-ZEPLIN:2022xrq}.
\begin{figure}
	\begin{center}
		\includegraphics[width=0.85\textwidth]{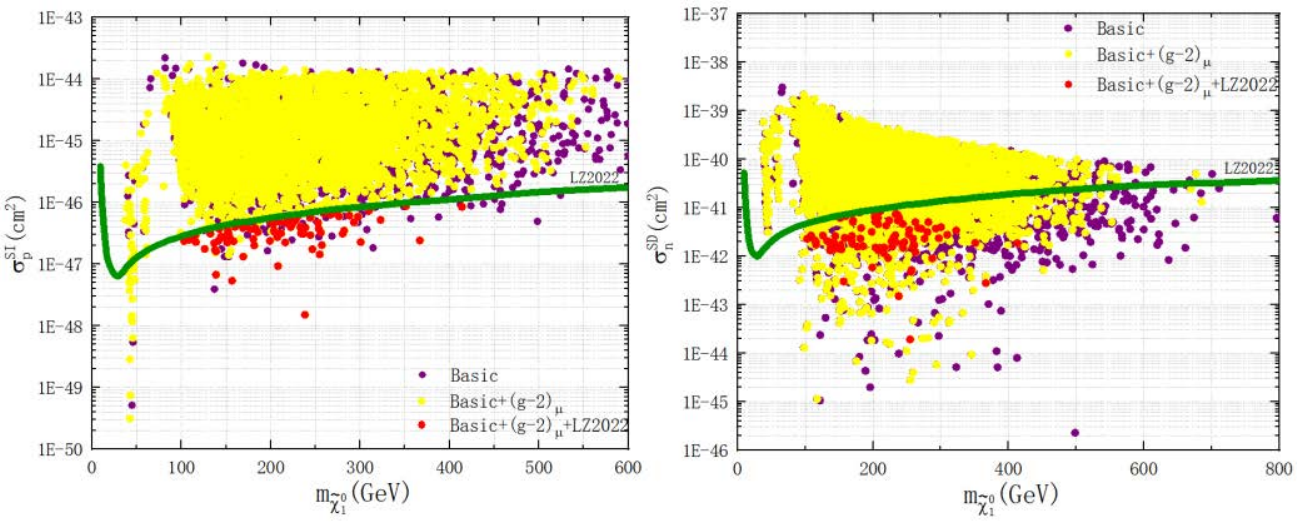}
	\end{center}
	\caption{ SI (left plot) and SD (right plot) nucleon-DM cross-section versus the
mass of DM. Purple samples satisfy the basic constraints; yellow samples satisfy the anomaly of $(g-2)_\mu$ within 2$\sigma$ level further, and red samples satisfy the basic constraints, muon anomaly constraint  within 2$\sigma$ level and also the LZ2022 experiment constraint.}
	\label{new-fig1}
\end{figure} 

Fig.\ref{new-fig1} shows the surviving samples on the plane $m_{\tilde{\chi}_1^0}-\sigma^{SI}_p$ and  $m_{\tilde{\chi}_1^0}-\sigma^{SD}_n$. The green line on the left (right) plot is the upper limit of SI (SD) nucleon-DM cross-sections, which comes from the results of the recent LZ2022 experiment. The samples above the green line are excluded
by the LZ2022 experiment. 
From this figure, we conclude that the results of recent nucleon-DM experiments impose
strong constraints on the parameter space in the $\mu$NMSSM, and the SD limit is complementary to the SI limit in limiting the parameter space in the $\mu$NMSSM. The figure also reveals
100 GeV $\lesssim m_{\tilde{\chi}_1^0} \lesssim $ 400 GeV considering the constraints from the recent LZ2022 experiment. 

\begin{figure}
	\begin{center}
		\includegraphics[width=0.85\textwidth]{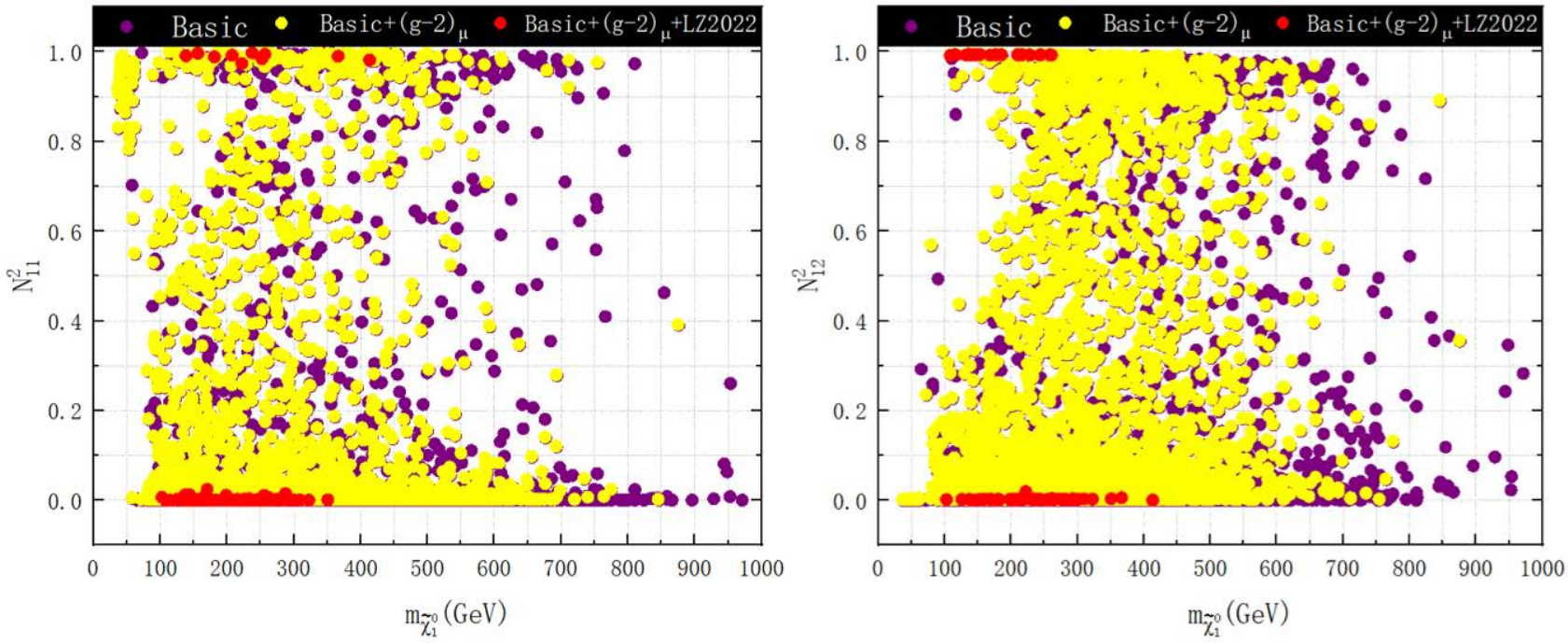}
  		\includegraphics[width=0.85\textwidth]{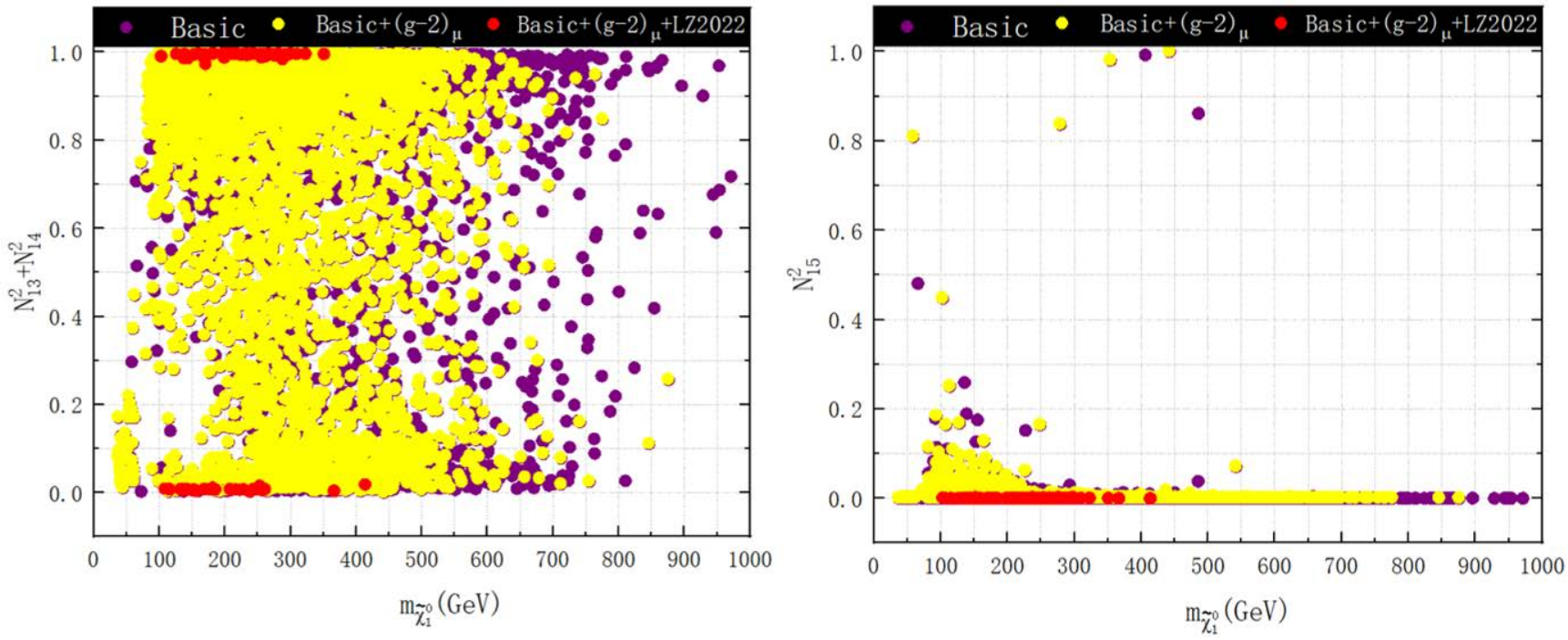}
	\end{center}
	\caption{Similar with Fig.\ref{new-fig1}, but shows the DM components versus the the mass of DM.		}
	\label{new-fig2}
\end{figure}  

\begin{figure}
	\begin{center}
		\includegraphics[width=1\textwidth]{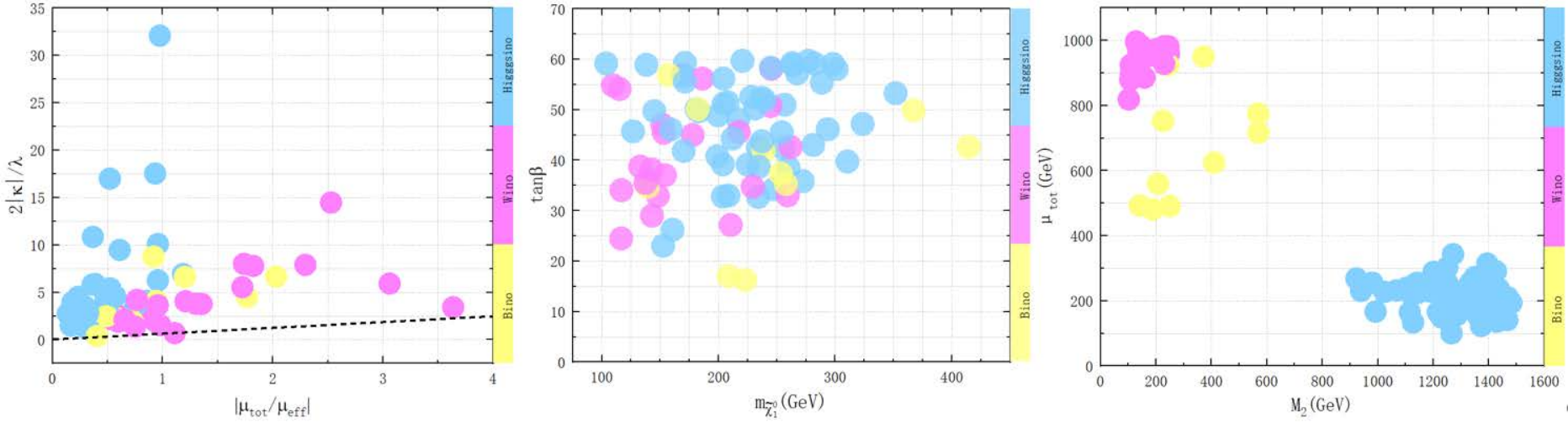}
	\end{center}
	\caption{Survived samples projected onto $|\mu_{tot}/\mu_{eff}|-2|\kappa|/\lambda$, $m_{\tilde{\chi}_1^0}-\tan{\beta}$ and
 $M_2-\mu_{tot}$ planes. The yellow points denote the Bino-dominated DM, light purple points denote the Wino-dominated DM, and blue points denote the Higgsino-dominated DM.}
	\label{new-fig3}
\end{figure} 
\begin{figure}
	\begin{center}		\includegraphics[width=0.85\textwidth]{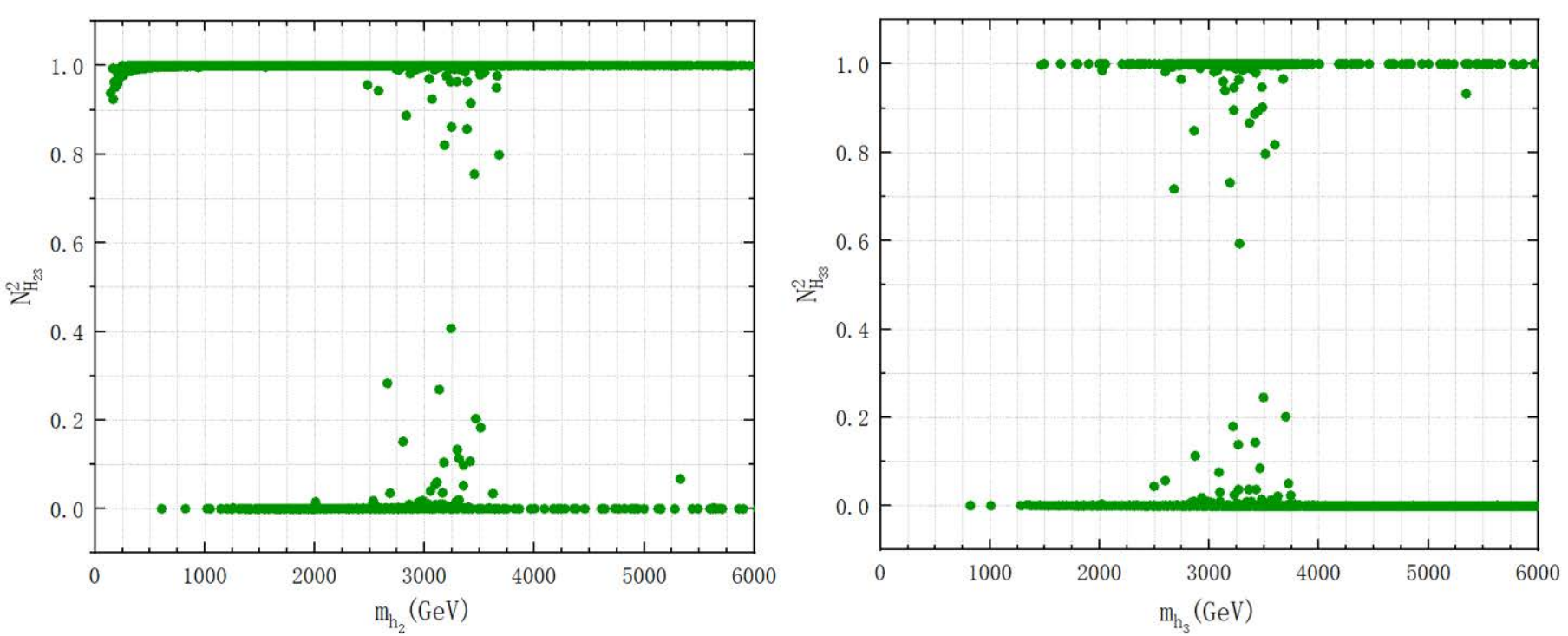}
    \includegraphics[width=0.85\textwidth]{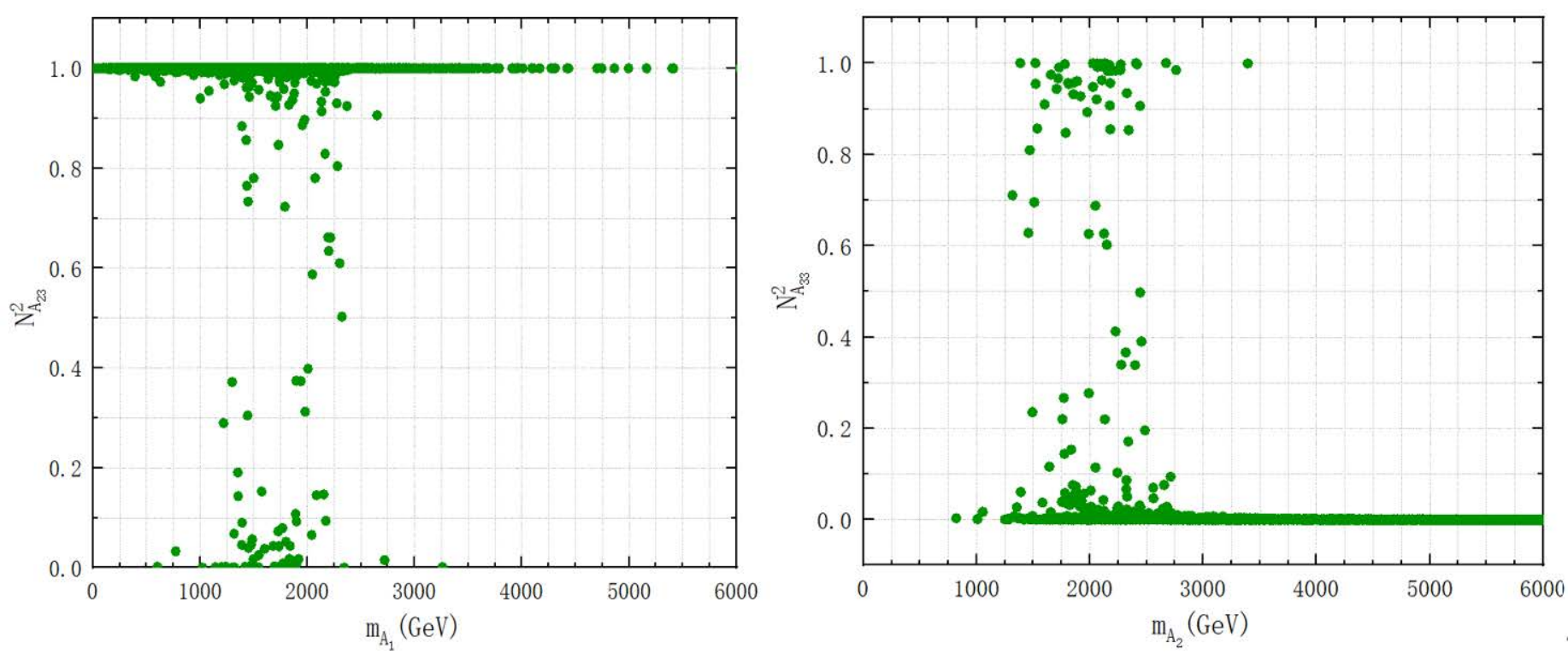}
	\end{center}
	\caption{The singlet component of the non-SM CP-even and CP-odd Higgs bosons versus their masses.}
	\label{higgscomponent}
\end{figure}   

We display the characteristics of DM components in Fig.\ref{new-fig2}. The upper left plot exhibits the Bino-component $N^2_{11}$, the upper right plot exhibits the Wino-component $N^2_{12}$, the lower left plot shows the Higgsino-component $N^2_{13}+N^2_{14}$, and the lower right plot shows the Singlino-component $N^2_{15}$.
Considering the constraints from the anomaly of $(g-2)_\mu$ and the recent LZ2022 experiment, the dark matter are mainly Wino-dominated or Higgsino-dominated. A few samples are Bino-dominated, but no samples are Singlino-dominated. The mass of Wino-dominated DM is less than 300 GeV, the mass of Higgsino-dominated DM is less than 350 GeV, and the mass of Bino-dominated DM is less than 400 GeV. 

To investigate the properties of the surviving parameter space, we pick out the red samples in Fig.\ref{new-fig1} and Fig.\ref{new-fig2} and project them onto 
$|\mu_{tot}/\mu_{eff}|-2|\kappa|/\lambda$, $m_{\tilde{\chi}_1^0}-\tan{\beta}$, $M_2-\mu_{tot}$ planes in Fig.\ref{new-fig3}.
For the Higgsino-dominated DM, 2$|\kappa|/\lambda$ is much larger than $|\mu_{tot}/\mu_{eff}|$ as can be seen from the left plot, which is significantly different from the scenario in the $Z_3$-NMSSM. In the $Z_3$-NMSSM, Higgsino-dominated DM only requires 2$|\kappa|/\lambda$ are larger than 1. 
The middle plot shows that $\tan{\beta}$ is greater than 
20 for the Wino-dominated or Higgsino-dominated DM. From the right plot, we can see that 
the survival samples mainly tend to be 850 GeV $ \lesssim M_2 \lesssim $ 1500 GeV and 
100 GeV $\lesssim \mu_{\text{tot}} \lesssim$ 300 GeV for Higgsino-dominated DM, and 
100 GeV $ \lesssim M_2 \lesssim $ 300 GeV and 800 GeV $\lesssim \mu_{\text{tot}} \lesssim$ 1000 GeV for the Wino-dominated DM.

\subsection{Properties of heavy Higgs bosons}
 
We pick out the survival samples satisfying
the basic constraints mentioned above, and also the constraints from the anomaly of 
$(g-2)_\mu$ and the limit of SI (SD) nucleon-DM cross-sections to investigate the properties of heavy Higgs bosons $h_3$ and $A_2$. 
In Fig.\ref{higgscomponent}, we show the singlet component of the non-SM CP-even and CP-odd Higgs bosons. From the upper plots, we can see that for most of the survival samples, the next-to-lightest CP-even Higgs boson $h_2$ can be mostly singlet-dominated or doublet-dominated. 
Correspondingly, the heaviest CP-even Higgs boson $h_3$ can be mostly doublet-dominated or singlet-dominated. But for a portion of the samples, singlet-doublet mixing can be large.
The lower plots show that, for most of the surviving samples, the lightest CP-odd Higgs boson $A_1$ is mostly singlet-dominated and the heaviest CP-odd Higgs boson $A_2$ is mostly 
doublet-dominated. However, for a part of the samples, singlet-doublet mixing can be large. And for a small fraction of the samples, $A_2$ can be mostly singlet-dominated.
 \begin{figure}
	\begin{center}
		\includegraphics[width=0.85\textwidth]{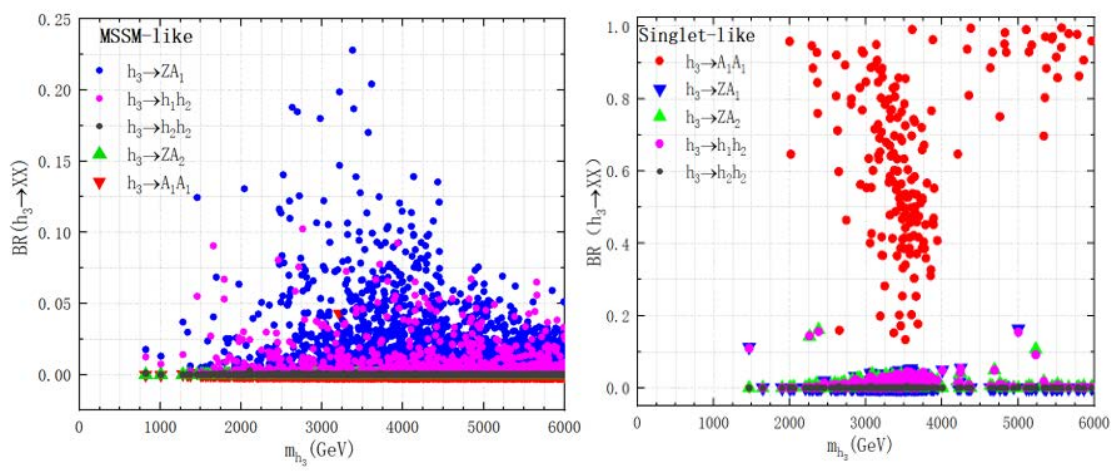}
	\end{center}
	\caption{The exotic decay channels of the heavy CP-even Higgs boson $h_3$. The left plot denotes $h_3$ being doublet-dominated (MSSM-like), and the right plot denotes $h_3$ being 
 singlet-dominated (singlet-like).}
	\label{h3decay}
\end{figure} 

 \begin{figure}
	\begin{center}
		\includegraphics[width=0.85\textwidth]{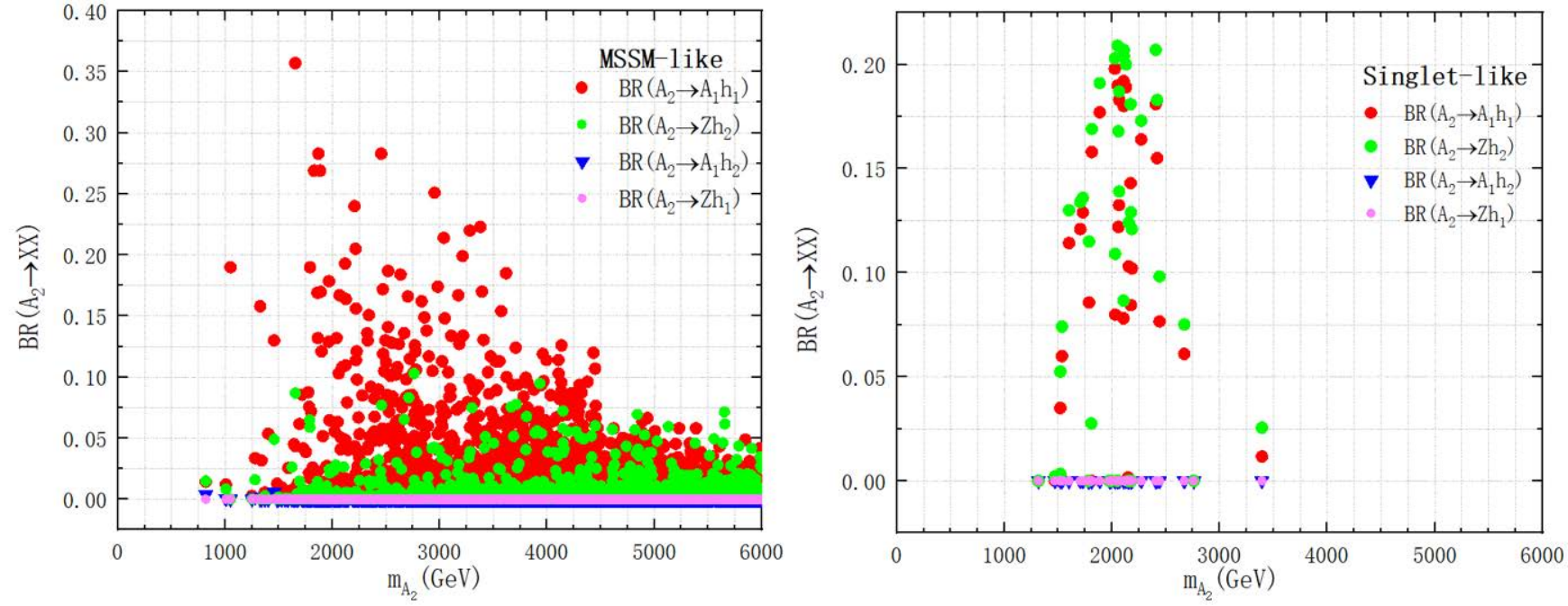}
	\end{center}
	\caption{The exotic decay channels of heavy CP-odd Higgs boson $A_2$. The left plot denotes $A_2$ being doublet-dominated (MSSM-like), and the right plot denotes $A_2$ being singlet-dominated (singlet-like).}
	\label{A2decay}
\end{figure} 

As discussed above, the exotic decay channels of heavy Higgs bosons $h_3$ and $A_2$ are open. In Fig.\ref{h3decay} and Fig.\ref{A2decay} we show the exotic decay channels of $h_3$ and $A_2$, and we only consider the decay channels of heavy Higgs boson decaying into lighter Higgs boson. The left (right) plot of Fig.\ref{h3decay} shows that $h_3$ is doublet-dominated (singlet-dominated), and the left (right) plot of Fig.\ref{A2decay} shows  $A_2$ is doublet-dominated (singlet-dominated). For the doublet-dominated Higgs boson $h_3$, 
the main decay channels are $h_3\rightarrow Z A_1$ and $h_3\rightarrow h_1 h_2$, and the 
branching ratio can reach about 23$\%$ and 10$\%$, respectively. The decay $h_3\rightarrow Z A_1$ is proportional to the  $A_{\rm NSM}$ component of the physical state $A_1$. The large branching ratio of $h_3\rightarrow Z A_1$ just corresponds to the scenario that the doublet component of $A_1$ is relatively large. The decay $h_3\rightarrow h_1 h_2$ is proportional to the Higgs trilinear coupling shown in the first equation of  Eq.(\ref{trilinear coupling}), which is usually relatively large 
when the mixing between doublet and singlet scalar fields is large. 
The singlet-dominated Higgs boson $h_3$ mainly decays into $A_1A_1$, and the branching ratio of $h_3\rightarrow A_1 A_1$ can reach to about 1.  
 
\begin{figure}
	\begin{center}		\includegraphics[width=0.85\textwidth]{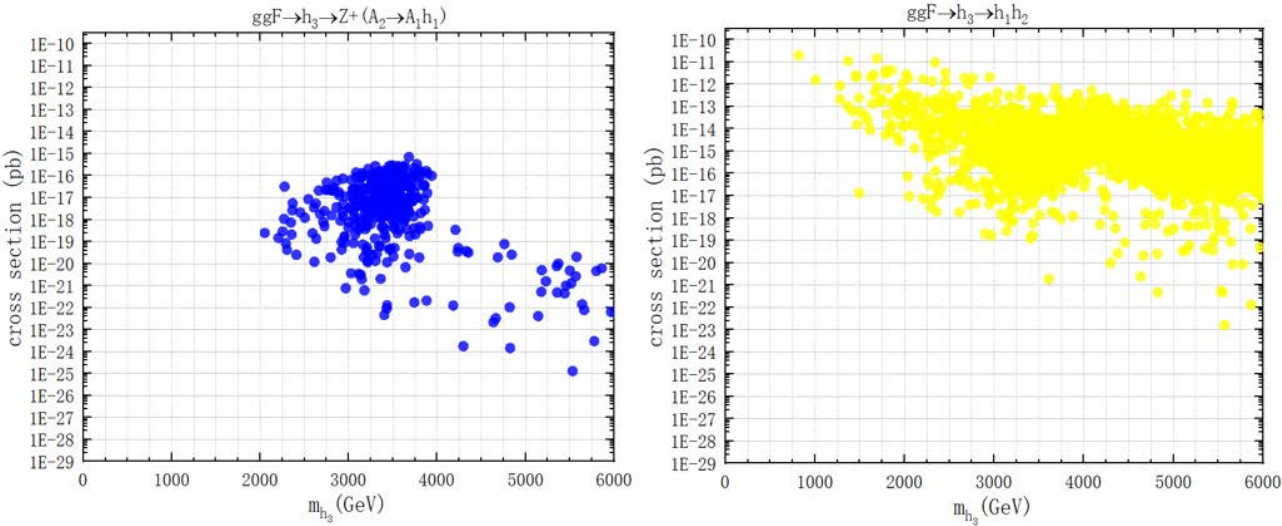}
	\end{center}
	\caption{The cross-section of the heavy CP-even Higgs boson $h_3$ decaying into the SM-like Higgs boson at 13TeV LHC.}
	\label{CP-even cross-section}
\end{figure} 

\begin{figure}
	\begin{center}
		\includegraphics[width=0.85\textwidth]{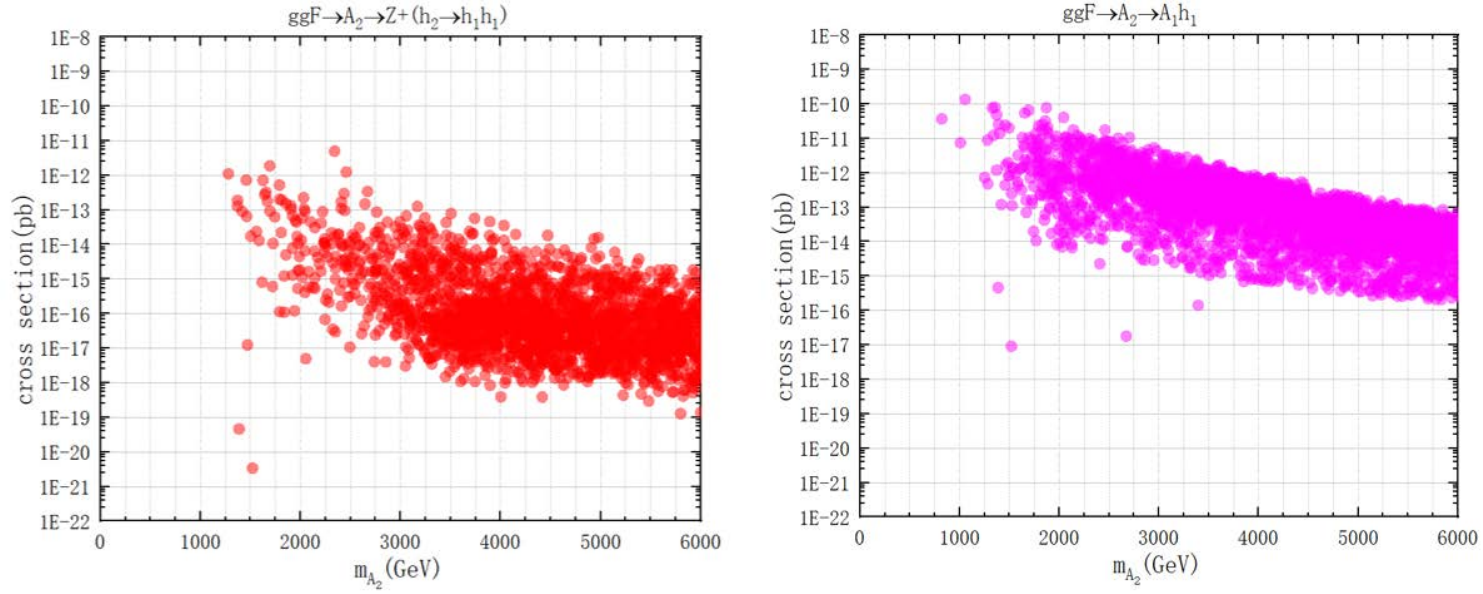}
	\end{center}
	\caption{The cross-section of the heavy CP-odd Higgs boson $A_2$ decaying into the SM-like Higgs boson at 13 TeV LHC.}
	\label{CP-odd cross-section}
\end{figure} 

Fig.\ref{A2decay} shows that, for the doublet-dominated Higgs boson $A_2$, 
the main decay channels of the Higgs boson $A_2$ are 
$A_2\rightarrow A_1 h_1$ and $A_2\rightarrow Z h_2$, and the branching ratio can reach to 
about 35$\%$ and 10$\%$, respectively. The decay $A_2\rightarrow A_1 h_1$ is proportional to the Higgs trilinear coupling shown in the second equation of Eq.(\ref{trilinear coupling}), 
which is usually relatively large when the mixing between doublet and singlet pseudoscalar fields is large, as the off-diagonal element $M^2_{P,12}$ shown. 
The decay $A_2\rightarrow Z h_2$ is proportional to the  $H_{\rm NSM}$ component of the physical state $h_2$. The large branching ratio of $A_2\rightarrow Z h_2$ just corresponds to the scenario that the doublet component of $h_2$ is relatively large. The branching ratio of
the decay $A_2\rightarrow Z h_1$ approaches to 0 because the $H_{\rm NSM}$ component of the SM-like $h_1$ is much lower.
The main decay channels of the singlet-dominated Higgs boson $A_2$ are 
$A_2\rightarrow A_1 h_1$ and $A_2\rightarrow Z h_2$.

Since the production cross-section of singlet-dominated Higgs bosons at the LHC is very small, we only
consider the production of doublet-dominated Higgs bosons $h_3$ and $A_2$. In
Fig.\ref{CP-even cross-section} and Fig.\ref{CP-odd cross-section}, we show the production
cross-section of the Higgs bosons $h_3$ and $A_2$ with $h_3$ and $A_2$  
decaying into the SM-like Higgs at $\sqrt{s}=13$ TeV LHC. We find that the processes $ggF\rightarrow h_3\rightarrow h_1 h_2$ and $ggF\rightarrow A_2\rightarrow  A_1 h_1$ are significant, and the cross-section can reach to about $10^{-11}$pb and $10^{-10}$pb,
respectively. 

\section{Summary}
In this paper, we have performed phenomenological studies on the properties of dark matter and heavy Higgs bosons $h_3$ and $A_2$ in the $\mu$NMSSM. Considering the basic constraints from Higgs data, DM relic density, and LHC searches for sparticles, we scan the parameter space of the $\mu$NMSSM.  We find that the LZ2022 experiment has a strict constraint on the parameter space of the $\mu$NMSSM, and the limits from the DM-nucleon SI and SD cross-sections are complementary. Samples surviving the LZ2022 experiment and the muon anomaly constraint at 2$\sigma$ level are mainly featured by $\tan{\beta} \lesssim 20$, 850 GeV $ \lesssim  M_2 \lesssim  $ 1500 GeV and 100 GeV $\lesssim  \mu_{\text{tot}} \lesssim $ 300 GeV for Higgsino-dominated DM, or 100 GeV $ \lesssim M_2 \lesssim $ 300 GeV and 800 GeV $\lesssim \mu_{\text{tot}} \lesssim$ 1000 GeV for Wino-dominated DM.

The detections of heavy Higgs bosons through the exotic decay modes into SM-like Higgs boson are important for analyzing the Higgs signals. We find that for doublet-dominated Higgs $h_3$ and $A_2$, the main exotic decay channels are $h_3\rightarrow Z A_1$, $h_3\rightarrow h_1 h_2$, $A_2\rightarrow A_1 h_1$ and $A_2\rightarrow Z h_2$, and the 
branching ratio can reach about 23$\%$, 10$\%$, 35$\%$ and 10$\%$ respectively.
At the 13TeV LHC, the production cross-section of processes $ggF\rightarrow h_3\rightarrow h_1 h_2$ and 
$ggF\rightarrow A_2\rightarrow  A_1 h_1$ can reach to about $10^{-11}$pb and $10^{-10}$pb,
respectively. 

\bibliographystyle{utphys}
\bibliography{ref.bib}
\end{document}